\documentclass[10.5pt,twocolumn,showpacs,showkeys,preprintnumbers,amsmath,amssymb,prb,superscriptaddress]{revtex4}
\usepackage[utf8]{inputenc}
\usepackage{amsmath}
\usepackage{amsfonts}
\usepackage{epsfig}
\usepackage{bm}
\usepackage{setspace}
\usepackage{booktabs} 
\usepackage{graphicx}
\usepackage[colorlinks=true, citecolor=red, linkcolor=blue, urlcolor=blue]{hyperref}
\begin{document}
\title{Structures and stability of the \href{https://youtu.be/j0Lw8NKYUYw}{Cu$_{38}$ cluster} at finite temperature}
\author{C\'esar Castillo-Quevedo}
\affiliation{Departamento de Fundamentos del Conocimiento, Centro Universitario del Norte, Universidad de Guadalajara, Carretera Federal No. 23, Km. 191, C.P. 46200, Colotl\'an, Jalisco, M\'exico}
\author{Edgar Paredes-Sotelo}
\affiliation{Departamento de Investigaci\'on en Pol\'imeros y Materiales, Edificio 3G. Universidad de Sonora. Hermosillo, Sonora, M\'exico}
\author{Carlos Emiliano Buelna-Garc\'ia}
\affiliation{Departamento de Investigaci\'on en Pol\'imeros y Materiales, Edificio 3G. Universidad de Sonora. Hermosillo, Sonora, M\'exico}
\author{Edwin Rene Hoil-Canul}
\affiliation{Universidad Polit\'ecnica de Tapachula, Carretera Tapachula a Puerto Madero km 24+300, San Benito, Puerto Madero C.P. 30830, Tapachula, Chiapas, M\'exico}
\author{Jhonny Robert Mis-May}
\affiliation{Universidad Polit\'ecnica de Tapachula, Carretera Tapachula a Puerto Madero km 24+300, San Benito, Puerto Madero C.P. 30830, Tapachula, Chiapas, M\'exico}
\author{Jarbin Barrios-D\'iaz}
\affiliation{Universidad Polit\'ecnica de Tapachula, Carretera Tapachula a Puerto Madero km 24+300, San Benito, Puerto Madero C.P. 30830, Tapachula, Chiapas, M\'exico}
\author{Martha Fabiola Martin-del-Campo-Solis}
\affiliation{Departamento de Fundamentos del Conocimiento, Centro Universitario del Norte, Universidad de Guadalajara, Carretera Federal No. 23, Km. 191, C.P. 46200, Colotl\'an, Jalisco, M\'exico}
\author{Edgar Zamora-Gonzalez}
\affiliation{Departamento de Fundamentos del Conocimiento, Centro Universitario del Norte, Universidad de Guadalajara, Carretera Federal No. 23, Km. 191, C.P. 46200, Colotl\'an, Jalisco, M\'exico}
\author{Adolfo L\'opez-S\'anchez}
\affiliation{Universidad Polit\'ecnica de Tapachula, Carretera Tapachula a Puerto Madero km 24+300, San Benito, Puerto Madero C.P. 30830, Tapachula, Chiapas, M\'exico}
\author{Jes\'us Ram\'on Cob-Cantu}
\affiliation{Universidad Polit\'ecnica de Tapachula, Carretera Tapachula a Puerto Madero km 24+300, San Benito, Puerto Madero C.P. 30830, Tapachula, Chiapas, M\'exico}
\author{Jorge Brice\~no-Mena}
\affiliation{Universidad Polit\'ecnica de Tapachula, Carretera Tapachula a Puerto Madero km 24+300, San Benito, Puerto Madero C.P. 30830, Tapachula, Chiapas, M\'exico}
\author{Freddy Francisco Agust\'in-Arg\"uello}
\affiliation{Universidad Polit\'ecnica de Tapachula, Carretera Tapachula a Puerto Madero km 24+300, San Benito, Puerto Madero C.P. 30830, Tapachula, Chiapas, M\'exico}
\author{Tzarara L\'opez-Luke}
\affiliation{Instituto de Investigaci\'on en Metalurgia y Materiales, Universidad Michoacana de San Nicol\'as de Hidalgo, Edificio U, Ciudad Universitaria, Morelia, Mich, 58030, M\'exico}
\author{Gerardo Mart\'inez-Guajardo}\email[email:]{germtzguajardo@uaz.edu.mx}
\affiliation{Unidad Acad\'emica de Ciencias Qu\'imicas, \'Area de Ciencias de la Salud, Universidad Aut\'onoma de Zacatecas, Km. 6 carretera Zacatecas-Guadalajara s/n, Ejido La Escondida C. P. 98160, Zacatecas, Zac.}
\author{Jose Luis Cabellos}\email[email:]{sollebac@gmail.com, jose.cabellos@uptapachula.edu.mx}
\affiliation{Universidad Polit\'ecnica de Tapachula, Carretera Tapachula a Puerto Madero km 24+300, San Benito, Puerto Madero C.P. 30830, Tapachula, Chiapas, M\'exico}
\date{\today}
\begin{abstract}

The UV-visible and IR properties of the \href{https://youtu.be/j0Lw8NKYUYw}{Cu$_{38}$} nanocluster depend to a great extent on the temperature. Density functional theory and nanothermodynamics can be combined to compute the geometrical optimization of isomers and their spectroscopic properties in an approximate manner. In this article, we investigate entropy-driven isomer distributions of \href{https://youtu.be/j0Lw8NKYUYw}{Cu$_{38}$} clusters and the effect of temperature on their UV-visible and IR spectra. An extensive, systematic global search is performed on the potential and free energy surfaces of \href{https://youtu.be/j0Lw8NKYUYw}{Cu$_{38}$}  using a two-stage strategy to identify the lowest-energy structure and its low-energy neighbors. The effects of temperature on the UV and IR spectra are considered via Boltzmann probability. The computed UV-visible and IR spectrum of each isomer is multiplied by its corresponding Boltzmann weight at finite temperature. Then, they are summed together to produce a final temperature-dependent, Boltzmann-weighted UV-visible and IR spectrum. Additionally,  Molecular Dynamics simulation of the \href{https://youtu.be/j0Lw8NKYUYw}{Cu$_{38}$} nanocluster was performed to gain insight into the system dynamics and make a three-dimensional movie of the system with atomistic resolution. Our results show the thermal populations at the absolute temperature of {Cu$_{38}$} cluster, and the disordered structure that dominates at high temperatures.
\end{abstract}

\maketitle
\section{Introduction}
Nano clusters are of interest due they allow us to study
the transition from free atoms to bulk condensed systems\cite{B517312B}
as a consequence, analyze the size-dependent
evolution of their properties.\cite{doi:10.1021/cr040090g}
Especially, Noble-Metal Nanoclusters (NMC) have
attracted attention in many fields of science due to
interesting plasmonic, catalytic properties,\cite{Inwati2018,https://doi.org/10.1002/ppsc.201400033} and
photophysical properties at nanoscale,\cite{Xavier2012}
Particularly, Nano Cu clusters embedded in the dielectric
matrix have attracted attention because of their tunable longitudinal surface
plasmon resonance.\cite{Inwati2018} Besides, copper is cheaper  than gold and silver,
and it  has large photosensitivity,  high thermal and electric conductivity, and
optical properties~\cite{Zhang2019} that makes it a good candidate to
develop nanodevices~\cite{C5RA14933A} and nanoelectronics.~\cite{Jena10560}
In particular, Cu$_{38}$ attracted attention due to it has   
a magic structures,~\cite{B912501A} defined in terms of geometric and
energetic factors and  related to the closing of electronic
shells~\cite{PhysRevB.69.235421} as it happens in small sodium clusters.\cite{RevModPhys.65.611}
For the Cu$_{38}$ cluster its magicity is due to only energetic considerations.\cite{A709249K,PhysRevB.69.235421}
In contrast, small packed barium clusters with magic numbers, the stability is
dominated for geometrical effects rather than electronic effects.\cite{RevModPhys.65.611}
It is believed  that magic structures are the putative global minimum energy structures 
on the potential energy surface, thus reflect the molecular
properties of the system.\cite{PhysRevB.69.235421}
From the experimental point of view, the Cu$_{38}$ cluster has been widely studied by
photoelectron spectroscopy techinque (PES). Pettiette et al.\cite{doi:10.1063/1.454575}
employing PES extract the electronic gap of anionic Cu$_{38}$ cluster and found
a semiconductor type with 0.33 eV.\cite{doi:10.1063/1.454575,Zhang2019} However, the
geometrical structure was not investigated. In contrast,  Kostko et al.
also studied the anionic Cu$_{38}$ and from the PES inferred  that putative global minimum should
be an oblate structure instead of a high symmetric structure,\cite{Kostko2005,Zhang2019} despite that computations for
38 atom clusters on noble metal clusters frequently found
high symmetric (cuboctahedral) structures.\cite{A709249K,Fujima1989,Kostko2005}
From the theoretical point of view, Taylor et al.
presented a study based on density functional theory (DFT) of
thermodynamic properties of Cu$_{38}$ cluster,\cite{Taylor_2008} Prevoius works
employing DFT studied the transition states and reaction energies of water gas shift reaction
in a Cu$_{38}$ cluster and Cu slab.\cite{doi:10.1142/S021963362050008X}
In other previously DFT studies the high symmetry octahedral structure was reported as the lowest
energy structure\cite{doi:10.1021/acs.jpcc.6b13086} employing PW91\cite{Vosko}
functional, plane wave basis set and pseudopotential approximation.\cite{doi:10.1063/1.3187934}
Hijazi et al.~\cite{Hijazi2010} investigated the Cu$_{38}$ cluster
employing hybrid strategy; they used the embedded atom method potential followed by DFT computations
using the PBE functional and pseudopotential approximation and reported 
that octahedral (OH) symmetry is the putative global minimum structure
followed by the incomplete-Mackay icosahedron (IMI) located at 0.26 eV
above of the putative global minimum.
On the other hand, search for the lowest energy structure employing many
body potentials gives a cuboctahedral structure.\cite{A709249K,Grigoryan2005}
Erkoc et al.~\cite{PhysRevA.60.3053} employing an empirical potential-energy function,
which contains two-body atomic interactions~\cite{Erkoc1994} found 
that the fivefold symmetry appears as putative global minimum in Cu$_{38}$ cluster.
In contrast, the cuboctahedron structure was reported as putative global minimum in a previous
studies\cite{PhysRevB.73.115415} employing empirical
\emph{many-body Gupta} and \emph{Sutton-Chen} potentials.
Nevertheless, there has been some discussion
which is the lowest energy structure, some previous works consider the Cu$_{38}$ octahedron cluster,
as the putative global minimum\cite{Hijazi2010,doi:10.1021/acs.jpcc.6b13086,doi:10.1063/1.3187934}, in contrast, several others found that Cu$_{38}$ cluster with the truncated octahedron geometry is
energetically  more stable than other
configurations\cite{ZHAO2017111,Hijazi2010,doi:10.1063/1.3187934,doi:10.1021/acs.jpcc.5b05023,Darby,jp1048088,doi:10.1080/08927022.2011.616502}. We point out that the energy computed
with different methods such as DFT, MP2, or CCSDT, just to mention few of them, 
yield different energetic ordering.\cite{Buelna,molecules26185710,Puente} In the case of DFT, the
functional and basis set employed, ZPE energy correction, or energy of dispersion among others can
interchange the putative global minimum. 

Moreover, practical molecular systems and materials  needs warm temperature~\cite{PhysRevLett.79.1337}, so the molecular
properties at temperatures T are dominated by Boltzmann distributions of
isomers,\cite{Buelna,MENDOZAWILSON2020112912,molecules26133953} therefore, their properties are   
statistical averages over the ensemble of conformations.\cite{Buelna} The structure corresponding to the
global minimum ceases to be the most likely at high T so other structures prevail.
Furthermore, in small Ag clusters, the temperature leads the transition from the initial FCC phase to other
structural modifications,\cite{Redel2015} so it promotes the changes of fases in materials. Interesting,
at different temperatures than zero, the molecular system minimizes the Gibbs free energy and maximizes the
entropy.\cite{Buelna,molecules26133953} 

Although the search of the global and local minima is useful in understanding
reactivities and catalytic efficiencies, but such studies most of the time neglect
temperature dependent entropic contributions to free energy when  increasing temperature.
Taking temperature into account requires dealing
with nanothermodyanamics.\cite{Buelna,molecules26133953,Hill,Li-Truhlar,Truhlar,Baletto,Grigoryan,Bixon} The thermodynamics of clusters have
been studied by a variety of tools,\cite{calvo,Wales925,Buelna,Li-Truhlar,Truhlar}
like molecular-dynamics simulations on boron clusters~\cite{Gerardo} and Cu$_{38}$ clusters.\cite{Zhang2019} The cluster properties depends strongly of
the structure, size, composition, and temperature, so the first step in order to understanding
molecular properties is the elucidation of the  lowest energy structure and its isomers close in energy;
\cite{molecules26133953,Buelna,Baletto,Darby,Ohno,Buelna2021a} This is a complicated task due to several factors.\cite{Buelna,molecules26133953} As second step for understanding cluster properties relies on their spectroscopy which
gives insight into its structure and it was proposed as a way of detecting structural transformations
into clusters. The influence of temperature on the spectroscopy has been computed before for a
variety of clusters, for instance,~\cite{Uzi,Buelna,molecules26133953} such in the present study, for the neutral Cu$_{38}$ cluster, we use the statistical formulation of thermodynamics or
nanothermodynamics\cite{Li-Truhlar,Buelna,Truhlar,molecules26133953}
to compute thermodynamics properties and define the putative global minimum at temperatures
diferent from zero, evaluated the relative populations among the isomers and
computed UV-Visible and IR spectra as a Boltzmann weighted spectrum sum of individual spectra.
Our findings show that at hot temperatures at amorphous structure strongly dominate the
putative global minimum whereas the truncated octahedron dominates at cold temperatures. 
The remainder of the manuscript is organized as follows: Section
2 gives the computational details and a brief overview of the theory and the algorithms used. The
results and discussion are presented in section 3; The putative global at room temperature
and relative population in ranging temperatures from 20 to 1500K, and  the IR spectra as a function of temperature.  Conclusions are given in Section 4.

\section{Theoretical Methods} 
\subsection{Method to Explore the Free Energy Surface and Computational Details}
At temperatures different of zero, the Gibbs free energy determines the lowest-energy structure,\cite{10.3389/fchem.2022.841964}
whereas at temperature zero the enthalpy determines the putative global minimum.\cite{Buelna,molecules26133953}
A simple analysis of the Gibbs free energy given by $\Delta G=\Delta H-\Delta ST$
deals to a conclussion, in order to minimize the Gibbs free energy we must to maximize
the entropy.\cite{Buelna,molecules26133953,10.3389/fchem.2020.00757} Front the theoretical point of view, and first of all, with the aim to understand  molecular properties,
at temperatures non-zero, we must know the lowest Gibbs free energy structure
or the largest entropy structure, and all structures closest in energy to the
lowest energy structure\cite{Buelna,molecules26133953} or all high-entropy structures closet in entropy to
the largest entropy structure. We must keep in mind  that the experiment
are performed at non-zero or finite temperatures. From a very general point of view, it has been shown that the validity of DFT can be extended to finite temperatures by the concept of ensemble DFT.\cite{10.3389/fchem.2019.00106}
The search of the global minimum in atomic clusters is complicated task
due to the number of possible combinations grows exponentially
with the number of atoms  leading to a combinatorial explosion problem
among others.\cite{Buelna,molecules26133953} Despite that is not an easy task, several algorithms to explore globally 
the potential/free energy surface coupled to a local optimizer generally of any
electronic structure package have been
successfully employed in a target so far, i.e. \emph{AIRSS} approach,\cite{Pickard_2011}
simulated annealing,\cite{kirkpatrick, metropolis, xiang, yang, vlachos, granville} kick methodology,\cite{Sudip,Cui,Vargas-Caamal2,Vargas-Caamal,Cui2,Vargas-Caamal2015,Florez,Ravell,Hadad,Saunders,Saunders2,Grande-Aztatzi} and  genetic algorithms~\cite{Guo,Dong,Mondal,Ravell,Grande-Aztatzi,Kessler,Alexandrova,Buelna} among many others.
Global structure searching at the DFT level  are computationally expansive to be applicable to intermediate
and large clusters size, as we mentioned early, the number of candidates increase exponentially.
In this work we used two-stage procedure to explore efficiently the potential energy surface,
in the first stage  we perform a global search  using a empirical methodology,
where Gupta interaction potential were used to describe the Cu-Cu interactions with
default parameters taken from references\cite{B204069G,PhysRevB.48.22} and coupled to
Basin Hopping global optimization algorithm implemented in \emph{Python} code and part of
global search of GALGOSON code.\cite{Buelna,molecules26133953}
At the second stage, all the lowest energy structures from the first stage are symmetrized, followed by 
a DFT optimization that were performed using the Gaussian suite code\cite{gauss} employing
two exchange-correlation functionals, B3PW91, PBE and  two
basis set, def2SVP and LANL2DZ, with and without taking into account D3 version of
Grimme’s dispersion corrections\cite{Grimme} as implemented in Gaussian 09 code.\cite{gauss} The
Becke’s hybrid three-parameter\cite{doi:10.1063/1.464913,PhysRevA.38.3098} exchange-correlation functional
in combination with the Perdew and Wang GGA functional PW91~\cite{PhysRevB.45.13244,Vosko}
is known as B3PW91 exchange-correlation functional. The B3PW91 has been employed in other studies of
reactivity in copper clusters with good performance,\cite{ doi:10.1021/acs.jpcc.5b05023} 
whereas, the PBE exchange-correlation functional~\cite{PhysRevLett.77.3865} has shown good performace
in thermochemistry properties.\cite{doi:10.1063/1.3691197}  Regarding
the LANL2DZ~\cite{schaefer1977methods} basis set, that was used in previous studies of 
computations of copper-based molecules properties produced result close to experimental
values.~\cite{Legge} Not too long ago, in a prevoius DFT studies, the def2-SVP~\cite{Weigend}
gave good results in the computations of Cu-metal ligand bond lengths.\cite{Niu}
The true minimun energy structures are validated  by the vibrational analysis without
imaginary frequency.
\subsection{Thermochemistry Properties}
All the thermodynamic properties of an ensemble of molecules can be derived from
molecular partition function\cite{Buelna,molecules26133953,Dzib,e20040218} 
so, the molecular partition function contains all thermodynamic information
 in a similar way that the quantum  wavefunction contains all the information about the system.\cite{Buelna,Dzib}
Previous theoretical studies used the partition function to compute thermodynamic properties of Cu$_n$ clusters
(n=2, 150) as a function of temperature and demonstrated that the  magic clusters is temperature dependent\cite{Grigoryan}
Zhen Hua-Li et al.~\cite{Truhlar} computed the thermodynamics of unsupported neutral Al$_n$ ($2 < n< 65$) particles
evaluating rovibrational partition functions, they reported that the dominant cluster depends on temperature.   
and gives and overview of recent progress on the nanothermodynamics of metal nanoparticles.\cite{Li-Truhlar}
Christopher Sutton et al.~\cite{10.3389/fchem.2020.00757} in framework of atomistic thermodynamics predict the behavior of
materials at realistic temperatures. Recently, Buelna-Garcia et al.~\cite{Buelna,molecules26133953} used the partition function to compute the temperature-dependent relative population and IR spectra of neutral Be$_4$B$_8$ and anionic  Be$_6$B$_{11}$
clusters, also Dzib et al.~\cite{Dzib} employed a similar procedure to compute the reaction rate constants. 
Other previous theoretical studies computed the temperature-dependent entropic contibutions on
[Fe(pmea)(NCS)2] complex.\cite{Brehm}
In this study, the temperature-dependent Gibbs free energy is computed employing the partition function 
Q given in Equation,~\ref{partition} under the rigid rotor, harmonic oscillator, Born-Oppenheimer, ideal gas, and a
particle-in-a-box approximations.  We have to underline that it must take into account the anharmonicity to compare theory with experiment.\cite{D1SC00621E} 
\begin{figure*}[ht!]
\begin{center}  
  \includegraphics[scale=0.50]{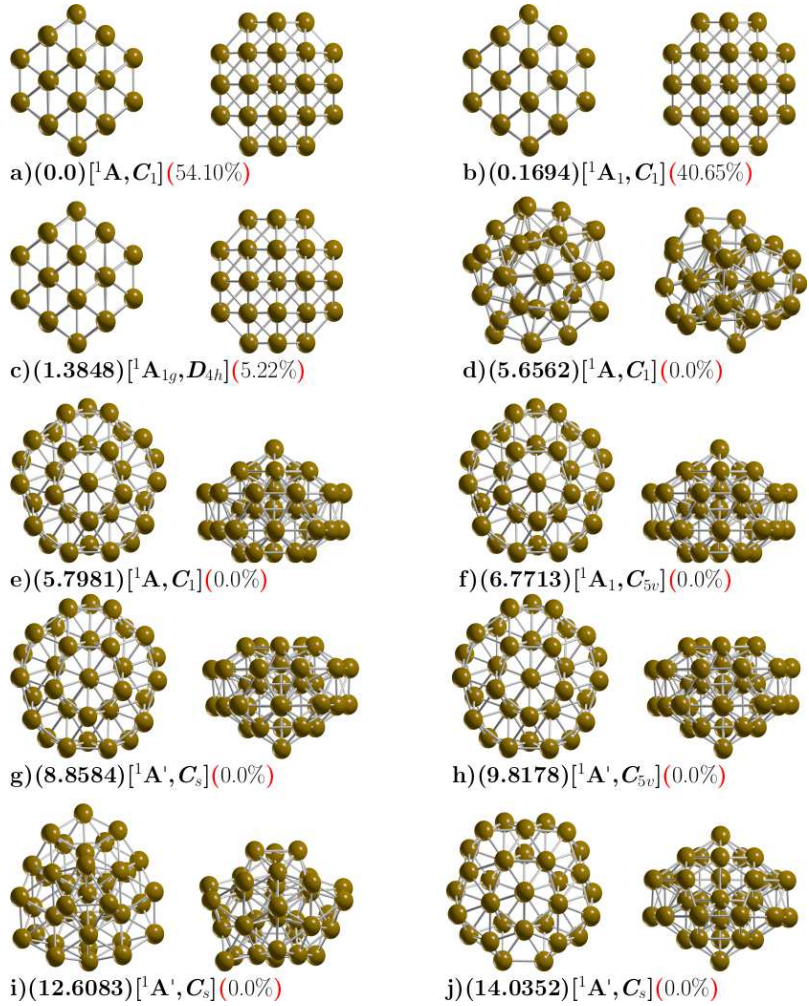}
\caption{(Color online) Optimized geometries in front and side
  views of neutral Cu$_{38}$ cluster at PBE-GD3/def2-SVP level of theory taking
  into account D3 version of Grimme’s dispersion
  corrections. The first letter is the isomer label,
  relative Gibbs free energies in kcal/mol (in round parenthesis) at 298.15 K,
  electronic group and group symmetry point [in square parenthesis], the probability of occurrence (in red round parenthesis) at 298.15 K, and the yellow-colored spheres  represent Cu atoms.} 
\label{geome}
\end{center}    
\end{figure*}
\begin{equation}
\displaystyle
Q(T)=\sum_{i}g_i~e^{{-\Delta{E_i}}/{K_BT}}
\label{partition}
\end{equation}
In Eq.~\ref{partition}, $g_i$ is the degeneracy factor, $k_{\textup{B}}$ is the Boltzmann constant, $T$ is the temperature, and ${-\Delta{E_i}}$ is the total energy of a cluster.\cite{Buelna,Dzib,mcquarrie1975statistical} We employ equations~\ref{e1} to~\ref{e4} to compute the internal energy (U), enthalpy (H), and Gibbs energy (G) of the Cu$_{38}$ cluster at finite temperature. The equations to compute  entropy contributions (S) are those employed in a previous work\cite{10.3389/fchem.2022.841964,Buelna,molecules26133953,Dzib} and any standard thermodynamics textbook.\cite{mcquarrie1975statistical,hill1986introduction}
\begin{equation}
\displaystyle
\mathcal{U}_0=\mathcal{E}_0+ZPE\\
\label{e1}
\end{equation}
\begin{equation}
\displaystyle
U_T=\mathcal{U}_0+(E_{ROT}+E_{TRANS}+E_{vib}+E_{elect})\\
\label{e2}
\end{equation}
\begin{equation}
\displaystyle
H=U_T+RT\\
\label{e3}
\end{equation}
\begin{equation}
\displaystyle
{G}=H-TS\\
\label{e4}
\end{equation}
In Equations above, ZPE is the zero-point energy correction  $\mathcal{E}_0$ is the electronic energy,
and $E_{ROT}+E_{TRANS}+E_{VIB}+E_{elect}$ are the contributions to energy due to
translation, rotation, vibration  and electronic as function of temperature, respectively.
To compute the Boltzmann probability of occurrence of one particular neutral Cu$_{38}$ cluster
in an ensemble at thermal equilibrium and at finite temperatures,   
we employ the probability of occurrence\cite{Buelna,molecules26133953,Truhlar,Bhattacharya,Bhumla,Shortle,MENDOZAWILSON2020112912,Dzib,Schebarchov,Goldsmith,Grigoryan,Truhlar} given in Equation~\ref{boltzman1} 
\begin{equation}
\centering 
\displaystyle
P_i(T)=\frac{e^{-\beta \Delta G^{k}}}{\sum e^{-\beta \Delta G^{k}}}\label{boltzman1},
\end{equation}
where $\beta=1/k_{\textup{B}}T$, and $k_{\textup{B}}$ is the Boltzmann constant, $T$ is the temperature, and $\Delta G^{k}$ is the Gibbs free energy of the $k^{th}$ isomer.  We point out that Gibbs free energies must be corrected considering the symmetry,  Buelna-Garcia et al.~\cite{molecules26133953} in a previous work shows that the contribution of the rotational entropy to the Gibbs free energy calculated with and without symmetry behave linearly with the temperature and could be significant~\cite{10.3389/fchem.2022.841964}
Equation~\ref{boltzman1} is restricted so that  the sum of all
probabilities of occurrence, at fixed temperature T, $P_i(T)$ is  equal to 1 
and given by Equation~\ref{bol2}
\begin{equation}
\centering 
\displaystyle
\sum_i P_i(T)=1\label{bol2},
\end{equation}

In this study, the Boltzmann  weighted UV-Vis spectrum  at finite temperature is given by Equation~\ref{vcd}
\begin{equation}
\displaystyle
UV-Vis=\sum_i^{n}(UV-Vis)_{i}\times P_i(T)
\label{vcd}
\end{equation}
Where $n$ is the total number of cluster in the ensemble, UV-Vis$_{i}$ is the UV-Vis of the $i^{th}$
isomer at temperature T=0, and  P$_i$(T) is the probability of occurrence of the $i$ isomer
given by Equation~\ref{boltzman1}. To compute the probability  of occurrence and the spectra 
we used the Boltzmann-Optics-Full-Ader code (\emph{BOFA}).\cite{Buelna,MENDOZAWILSON2020112912}
\section{Results and Discussion}
\begin{table*}[!htbp] \centering
\caption{The energetic isomer ordering employing B3PW91/Def2TZVP,  PBE/Def2TZVP, and PBE/LANL2DZ levels of theory.}
 \label{tab23}
 \begin{tabular}{@{\extracolsep{0.2pt}} ll l l lll lll l} 
 \\[-1.8ex]\hline 
 \hline \\[-1.8ex] 
  &   &  \multicolumn{9}{c}{\Large{Isomers (energy kcal/mol)}}\\
\cline{3-11}\\ [-1.8ex] 
 \multicolumn{1}{c}{\Large{Level of theory}} & \multicolumn{1}{l}{Energy} & \multicolumn{1}{c}{$i_a$} & \multicolumn{1}{c}{$i_b$ } & \multicolumn{1}{c}{$i_c$}& \multicolumn{1}{c}{$i_d$}& \multicolumn{1}{c}{$i_e$}  &\multicolumn{1}{c}{$i_f$} &\multicolumn{1}{c}{$i_g$}&\multicolumn{1}{c}{$i_h$} &\multicolumn{1}{c}{$i_i$}  \\
 \hline \\[-1.8ex] 
                             & $\Delta G$                                      &  0.0  &  0.16  & 1.38  & 5.65  & 5.79  & 5.81 &  6.76 & 8.85 & 9.81\\
   \large{B3PW91/Def2TZVP/GD3}   & \large{$\mathcal{E}_0+\mathcal{E}_{\mathrm{ZPE}}$} &  0.0  &  0.09  & 0.0   & 5.01  & 5.01  & 5.01 &  5.01 & 5.01 & 8.17\\
                             & \large{$\mathcal{E}_0$}                         &  0.05  &  0.0  & 0.10  & 8.76  & 4.89  & 4.89 &  4.88 & 4.88 & 4.88 \\
                   
 \hline \\[-1.8ex] 
                                 &  $\Delta G$                                      &  0.0  &  0.95  & 2.0   & 2.40  & 2.73  & 2.91  & 2.94 & 3.28 & 3.32 \\
\large{B3PW91/Def2TZVP    }     &  \large{$\mathcal{E}_0+\mathcal{E}_{\mathrm{ZPE}}$} &  0.0  &  0.0   & 2.14  & 3.29  & 6.52  & 3.84  & 2.14 & 3.27 & 3.84 \\
                                &  \large{$\mathcal{E}_0$}                         &  0.0  &  0.0   & 2.09  & 3.27  & 6.13  & 3.76  & 2.09 & 3.28 & 3.76 \\
  \hline \\[-1.8ex]
                             &  $\Delta G$                                          &  0.0  &  0.86  &  0.92 & 5.02  & 7.23  & 7.59  & 7.81 & 8.88 &  12.27     \\
 \large{PBE/Def2TZVP/GD3}        &  \large{$\mathcal{E}_0+\mathcal{E}_{\mathrm{ZPE}}$} &  0.0  &  0.89  &  0.0  & 8.70  & 7.94  & 7.92  & 8.61 & 7.94 &  14.24     \\
                             &  \large{$\mathcal{E}_0$}                             &  0.0  &  0.90  &  0.0  & 9.14  & 7.92  & 7.93  & 8.84 & 7.94 &  12.47     \\
  \hline \\[-1.8ex]
                             &  $\Delta G$                                      &  0.0  &  0.34  &  0.92   & 1.37  & 1.38  & 1.77  &2.82 & 5.79 & 8.70     \\
 \large{PBE/Def2TZVP}        &  \large{$\mathcal{E}_0+\mathcal{E}_{\mathrm{ZPE}}$} &  0.0  &  0.0   &  0.37  &  0.41  & 1.77  & 1.77  &1.77 & 9.08 & 15.75     \\
                             &  \large{$\mathcal{E}_0$}                         &  0.0  &  0.0   &  0.38  &  0.39  & 1.73  & 1.73  &1.74 & 9.47 & 9.86     \\
  \hline \\[-1.8ex]
                             &  $\Delta G$                                      &  0.0   &  1.02  &  3.37   & 3.46  & 8.63  & 9.11  &9.59 & 9.62 & 9.74     \\
  \large{PBE/LANL2DZ/GD3}   &  \large{$\mathcal{E}_0+\mathcal{E}_{\mathrm{ZPE}}$}  &  2.03  &  0.0   &  2.12   & 2.14  & 8.31  & 8.77  &8.73 & 9.34 & 8.71     \\
                             &  \large{$\mathcal{E}_0$}                         &  1.97  &  0.0   &  2.01   & 2.01  & 8.08  & 8.60  &8.52 & 9.12 & 8.45     \\
  \hline \\[-1.8ex]
                             &  $\Delta G$                                      &  0.0   &  2.15  &  2.87   & 3.03  & 3.26  & 4.31  & 8.89 & 9.66 & 9.85     \\
  \large{PBE/LANL2DZ}        &  \large{$\mathcal{E}_0+\mathcal{E}_{\mathrm{ZPE}}$} &  0.97  &  0.0   &  1.46   & 1.38  & 1.64  & 1.68  & 7.98 & 9.19 & 9.58     \\
                             &  \large{$\mathcal{E}_0$}                         &  0.86  &  0.0   &  1.27   & 1.11  & 1.48  & 1.42  & 8.03 & 8.99 & 9.37     \\
  \hline \\[-1.8ex]
\end{tabular}
\end{table*}
\subsection{The lowest energy structures and energetics}
The ball and stick model shown in Figure~\ref{geome} depicted
the lowest-energy structure and the low-energy structures of neutral Cu$_{38}$
clusters along with some competing isomers. At B3PW91/def2SVP level of theory and taking into
account the dispersion pairwise correction of Grimme (DFT-GD3),\cite{Grimme} ZPE energy corrections and
at room temperature and 1 atmospheric pressure. 
We found a tetrakaidecahedron as the
lowest energy structure which has 14 faces: six equivalent square fcc(100) and eight equivalent hexagons,
this shape is obtained when cutting the corners off 3D diamond  shape, and it is a fcc-like truncated
octahedron (TO). The calculated structure belongs to point group symmetry C$_1$, electronic ground state $^{1}$A,
its lowest IR active vibration frequency  is 32.57 cm$^{-1}$ and is a semiconductor with
electronic gap 0.623 eV. It is known that the bulk rare gas crystals have a face
centered cubic (FCC) crystalline symmetry.

Previous works on exploration of the potential energy surface of Cu$_{38}$ cluster
using  genetic algorithms with  Gupta potential  often find highly symmetric TO structure,\cite{Kostko2005}
other reported previous work employing \emph{Sutton-Chen} potential  with monte Carlo simulation also
find TO structure.\cite{Darby}
The optimized Cu-Cu bond length is found to be 2.4670,~\AA~  which is in good agreement with other
reported DFT calculations Cu-Cu dimmer\cite{PhysRevA.69.043203,PhysRevB.73.155436} of 2.248,~\AA~ and
is consistent with the experimental bonding distance Cu-Cu 2.22\AA.~\cite{PhysRevA.69.043203}
Our computed diameter of TO structure is 7.8~\AA~ and also is in good agreement of 8~\AA~ reported in
previous theoretical DFT calculations.\cite{PhysRevB.73.155436}
The second structure higher in energy  lies at 0.16 kcal/mol at temperature of 298.15 K also is a TO
structure with point group symmetry C$_1$, electronic ground state $^1$A,
the lowest IR active vibration frequency  is 32.13 cm$^{-1}$, and is a semiconductor with electronic
gap 0.623 eV, similarly to that of the putative global minimum. The next structure is slightly higher in energy located at 1.38 kcal/mol  also is a TO structure, but with D$_{4h}$ point group symmetry and
electronic ground state $^1$A$_{1g}$, and the lowest IR active vibration frequency  is 33.44 cm$^{-1}$.
We also explored TO structure, initializing the geometry from the high-symmetries OH and TH and after
geometry optimization without constrains, the OH and TH symmetries  become C$_1$ and D$_{H4}$ symmetries.
The perfect OH symmetry it could be deformed due to the \emph{Jahn-Teller effect}\cite{PhysRevB.73.155436,PhysRevA.69.043203} and would have to be taken into account
in the calculation of total energy\cite{PhysRevB.97.165204,ZLATAR201086}
and the relative population at temperature T could change as consequence in the
optical properties.~\cite{Opik}  Recently, in one of our previous works, we clarify the origin of
Gibbs free energy differences between two similar structures just with different group symmetry,
that is due to the rotational entropy, specifically the $R~T~ln(s)$ factor.\cite{molecules26133953}  In this work, the
energy difference of 0.16 between the two isomers depicted in Figure~\ref{geome}a with symmetry C$_1$ and
RMSD, the difference is 0.08 and is due to the Jean teller effect. The structure located at 1.38 kcal/mol
above the putative global minimum with D$_{4H}$ symmetry is due to rotational entropy. 
\begin{figure*}[ht!]
\begin{center}  
\includegraphics[scale=0.60]{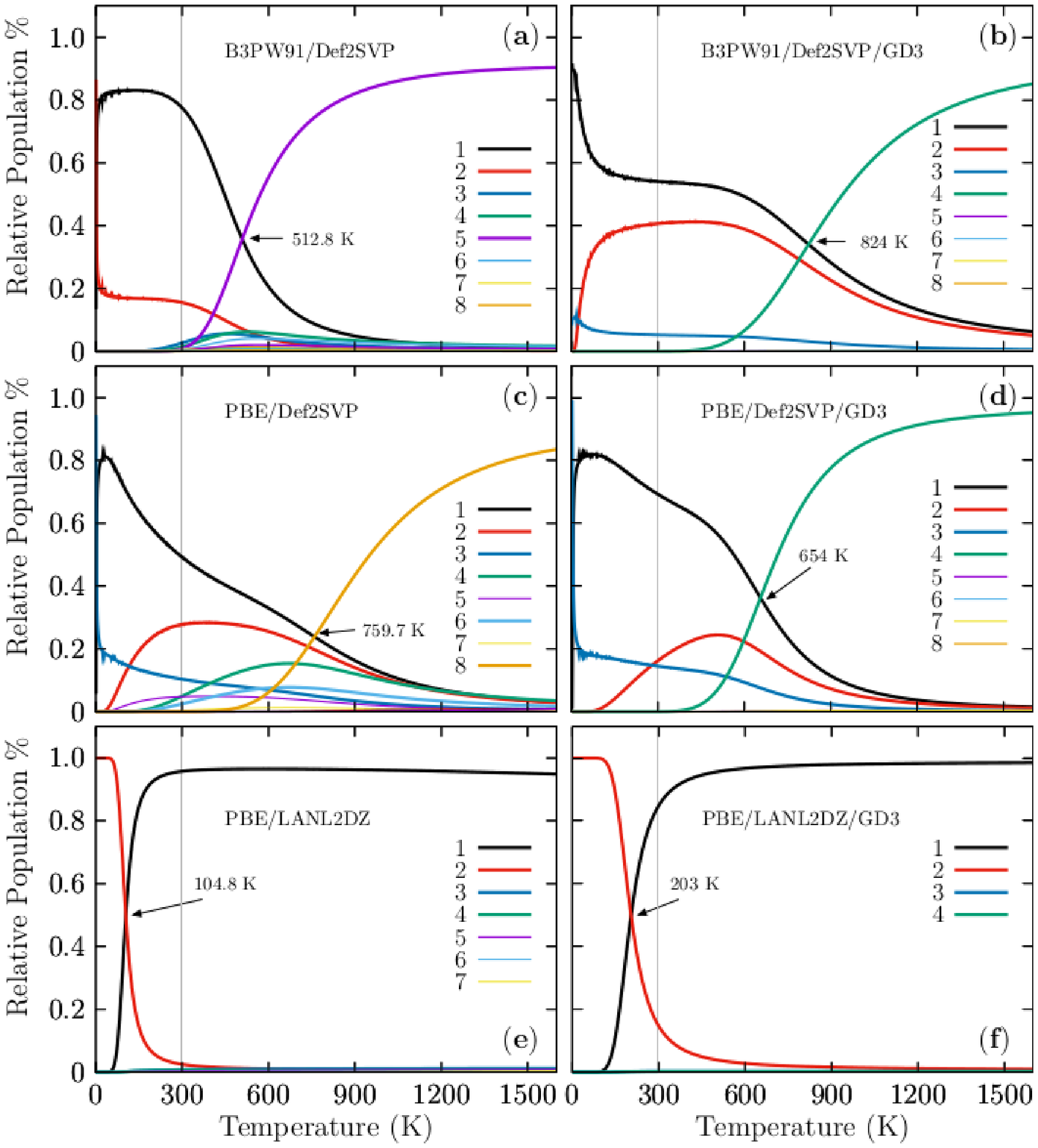}
\caption{(Color online) The relative stability/population or probabilities of occurrence
  for temperatures ranging from 20 to 1500 K at three different levels of theory
  B3PW91/def2-SVP, PBE/def2-SVP and PBE/LANL2DZ with and without D3 Grimme dispersion.
  The effect of dispersion on the solid-solid transformation point in temperature scale and
  in all cases is large for Cu$_{38}$ cluster.  At hot temperatures, in all cases
  the dominant structure is a amorphous geometry depicted in Figure~\ref{geome}(4)
  whereas at cold temperatures, the dominant structure can change from TO to
  inverted incomplete-Mackay icosahedron (IIMI) structures.  
   The TO structure depicted
   in Figure~\ref{geome}(1) is the strong dominant structure at cold temperatures
   at B3PW91-GD3/def2-SVP level of theory.} 
\label{popu}
\end{center}
\end{figure*}
The next structure, shown in Figure~\ref{geome}d is located at 5.65 kcal/mol above the putative global minimum,  with point group symmetry  C$_1$,  and  electronic ground state $^1$A,  the lowest IR active vibration frequency  is 24.16 cm$^{-1}$,  it is a distorted structure semiconductor with electronic gap of 1.0 eV, the calculated Cu-Cu bond distance is 2.50,~\AA~ and the molecular diameter of this structure is 9.1 kcal/mol, slightly larger  Cu-Cu bond distance and diameter than the global minimum, this structure possesses the the smallest relative ZPE energy, as shown in Figure, as well as the smallest  frequency of the vibrational modes of all isomers.
\begin{figure*}[ht!]
\begin{center}  
\includegraphics[scale=0.5]{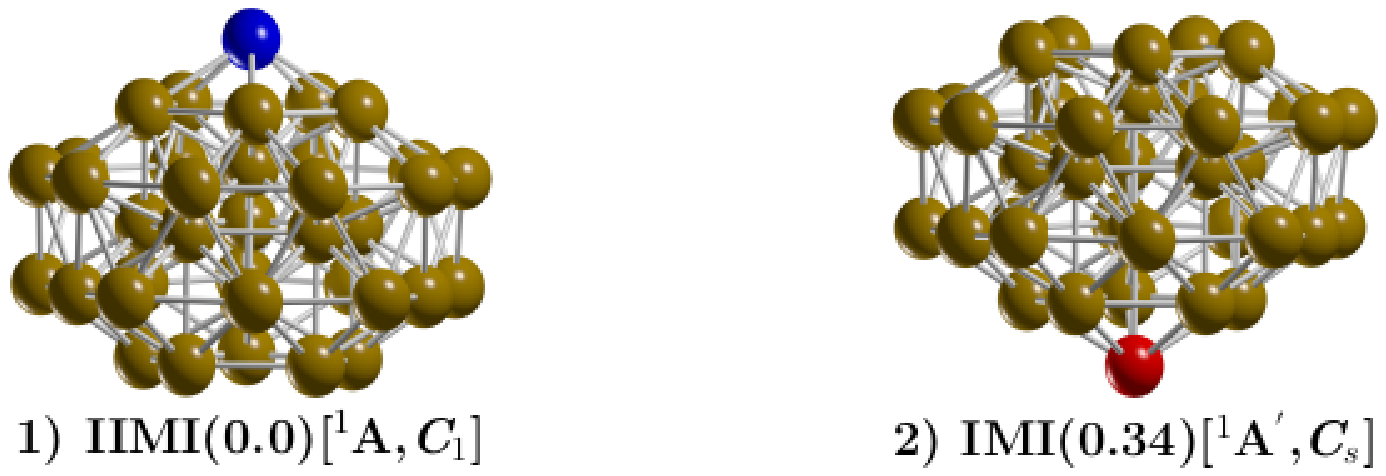}
\caption{We depicted the inverted incomplete-Mackay icosahedron (IIMI) label 1 with symmetry C$_1$
  and the incomplete-Mackay icosahedron (IMI) label 2 with symmetry C$_s$ and located 0.34 kcal/mol energy
  above the putative minimum global at 298.15 K. The yellow,
  red and blue colored spheres represent copper atoms. The IIMI structure is the result of interchanging
  the red Cu atom depicted in the IMI structure to the position of blue atom in the IIMI structure.
  The IMI structure was reported in reference\cite{Zhang2019} as the low-energy structure.
  The HOMO-LUMO gap of the IMI structure is 0.24 eV (0.356 eV reported by previous DFT studies\cite{Zhang2019})  whereas for the HOMO-LUMO gap for IIMI structure is 0.30 eV, suggesting why the IIMI structure is
  energetically more stable.}
\label{iimi}
\end{center}
\end{figure*}
The next two higher energy structures shows in   Figure~\ref{geome}(e,f) are located up to 5.8 kcal/mol, an both of them are the incomplete-Mackay icosahedron (IMI) with C$_1$ and C$_{5V}$ point group symmetries and
electronic ground state $^1$A and  $^1$A$_1$  respectively. For both cases,
the molecular diameter is 8.54,~\AA~ electronic gap of 0.97 eV eV  and the Cu-Cu bonding
distance is 2.47~\AA.~ There are other higher energy structures shows in
Figure~\ref{geome}(g,h,i,j), these do not contribute to any molecular property in
all temperature range.
In the Supplementary Information, Figure~\ref{fig.geo}, Appendix~\ref{appendix:d}, we depicted the
lowest energy structures screening at B3PW91/def2SVP level of theory and
without taking into account the atom-pairwise correction of Grimme GD3.
The lowest energy structure is IMI structure  with point group symmetry  C$_1$
and electronic ground state $^1$AH. The molecular diameter is 8.69~\AA,~slightly larger than
those found at TO structure of 7.8,~\AA~the average bond distance is 2.50~\AA.~At PBE-GD3/Def2SVP
level of theory, we found a IMI
structure to be the most stable structure whereas at the PBE-GD3/LANL2DZ level of theory
we found  the TO structure as the putative global minimum energy.  
The complete description of the structures located at higher energies is in the
Supplementary Information. We point out here,  for the Cu$_{38}$ clusters,
the order energetically of the isomers and the energy gap among isomers as well as the putative global minimum interchange when we take into account the 
dispersion interactions. The atomic XYZ coordinates (at B3PW91/def2SVP level of theory) are displayed in Appendix~\ref{appendix:f}
\subsection{Energetics}
In the computation of energies employing different methods yield different results due to
the functional and basis sets and therefore the energetic
isomer ordering changes.\cite{molecules26133953,Puente}
The comparison of two different exchange-correlation functionals and two basis set
and with and without taking into acount the dispersion D3 the Grimme 
are shown in the Table~\ref{tab23}. 
The optimizations performed at the B3PW91/PBE-def2TZVP considering the dispersion
yield the same type of lowest energy equilibrium geometry and similar energetic
isomer ordering when we employ the electronic energy with and without ZPE correction energy
and Gibbs free energy computed at room temperature.  
From the energetic point of view, the inclusion of dispersion is more important than
the type of functional and basis set, i.e The first line of the Table~\ref{tab23}
show the relative Gibbs free energies computed  at the B3PW91-D3/def2TZVP level of theory, the isomer
label i$_b$ in the Table~\ref{tab23} and depicted in Figure~\ref{geome}b is located
0.16 kcal/mol above the putative global minimum whereas the second line of the Table~\ref{tab23}
show the relative Gibbs free energies computed  at the B3PW91/def2TZVP level of theory and  the
isomer i$_b$ in the Table~\ref{tab23} and depicted in Figure~\ref{geome}b is located
0.95 kcal/mol above the putative global minimum
as it is shown in Table~\ref{tab23}. For isomer i$_b$, the effect of the dispersion
on the energy is decrease the relative Gibbs free energy with respect to the putative
global minimum. (from 0.95 to 0.16 kcal/mol). The effect of dispersion, in the case of
isomers i$_c$, is also decrease the relative Gibbs free energy with respect to the putative
global minimum from 2.0 to 1.38 kcal/mol whereas for isomer  label i$_d$, the relative Gibbs free energy
increase from 2.4 to 5.65 kcal/mol. In summary, the effect of dispersion reduces the Gibbs free energy in the
lowest energy structures where the Boltzamnn factors are difrentes from zero.
A overall comparison of free energies computed with  functional B3PW91, second line of the Table~\ref{tab23}, and PBE in four line
in the Table~\ref{tab23}, versus  free energies computed with  functional PBE, four line of the Table~\ref{tab23}, and PBE in four line in the Table~\ref{tab23}, shows a reduction in the relative Gibbs free energies
when the PBE functional is employed. For the case of the basis set the LANL2DZ
increase the relative Gibbs free energy in the low energy isomers as it is shows a comparison of
line 6 of Table~\ref{tab23} versus line 1 of Table~\ref{tab23}. 
\subsection{Relative stability}
The probability of finding the TO structure with C$_1$ symmetry at
B3PW91/def2-SVP level of theory is depicted in black-solid line in panel (a) of
Figure~\ref{geome}(4). It strongly dominates from 0 to 300 K, thus
all the molecular properties are due only to this structure. From slightly before 300 K,
it start to decay exponentially and almost disappear at 900 K, at the same time,
the  probability of finding the amorphous structure with point group symmetry C$_1$, is
depicted in violet-solid line in Figure~\ref{popu}(a), it start to grow exponentially
and become dominate at temperature above 512.8 K and at 900 K it
become strongly dominate. At solid-solid transition temperature of 512.8 K the TO and the amorphous
structure co-exist. The effect of dispersion can be appreciated in
panel (b) of Figure~\ref{popu}. The relative population is computed at
B3PW91-G3/def2-SVP level of theory. The effect of the dispersion is dramatic,
the solid-solid transformation point is shifted from 512.8 to 824 K, an increase of 160{\%}
and from the panel (b) one can see that the molecular properties below 600 K, are due to
only the TO structure. Slightly before 600 K, The probability of finding the amorphous
structure, depicted in green-solid line in panel (b),  start to grow up exponentially
and at temperature of 824 K the TO and amorphous structure co-exit. Whereas a temperature of 600
K the probability of find the TO structure start do decay exponentially and at 900 K its
value still around 20{\%}.
The probabilities of occurrence at the PBE/def2-SVP level of theory of a particular
Cu$_{38}$ isomer are displayed in panel (c) of Figure~\ref{popu}. The dominant putative
global minimum structure  at T=0, is the inverted incomplete-Mackay icosahedron (IIMI)
structure depicted in Figure~\ref{iimi}a with C$_1$ symmetry and its probrobability of finding it
is depicted as a black solid line in panel (c) of Figure~\ref{popu}. Its probability
decays almost linearly until 1000 K where it almost disappears. At temperature of
759.7 K, the solid-solid transformation point, the IIMI structure co-exist with
an amorphous structure. The probability of finding
the amorphous structure start to grew up at 600 K, and above of solid-solid
transformation point, it start to strongly dominate as putative global minimum. Where
the probability of finding the IMI structure is depicted in red-solid line  
on panel (c) of Figure~\ref{popu}, the largest probability is 30{\%} at
room temperature. Interesting, the probabilities of IIMI and IMI structures
does not cross at cold temperature. 
Zhang et al.~\cite{Zhang2019} reported that IMI structure could be highly competitive at finite temperatures,
but our findings shows that the amorphous structure with C$_1$ symmetry
is highly dominant at hot temperatures, whereas the IIMI structure is highly dominant at
cold temperatures. 

For ease comparison,  the Figure~\ref{iimi} display the IIMI and the IMI
structures side by side. At cold and the IIMI structure dominate.
The effects of dispersion is shift the solid-solid transformation point to lower temperatures
from 759.7 K to 654 K as one can see in panel (d) Figure~\ref{popu}.
The probability of finding the IIMI structure as a function of temperature is
depicted in black solid-line and it decays approximately linearly from 50 to 500 K, after that
it decays exponentially until 900 K where it disappears. At around 400 K, the probability of finding
the amorphous structure, depicted in green solid-line in panel (d) of Figure~\ref{popu},
start to grow up in exponential way, and at temperature of 654 K it co-exit
with the IIMI structure. Above 654 K, the amorphous structure becomes energetically favorable.
\subsection{IR spectra at finite temperature}
The properties observed in a molecule are statistical averages over the ensemble of geometrical conformations or isomers accessible to the cluster,
so the molecular properties are governed by the Boltzmann distributions of isomers that can
change significantly with the temperature primary due to entropic effects.\cite{molecules26133953,Buelna,Truhlar} The major contributions to the entropy are the many soft vibrational modes that the clusters possesses.
The IR spectrum is related to vibrations or rotations that alter the dipole moment, and it will happen in molecules that have a dipole moment. Also, the IR spectrum is related to the curvature of the potential curve versus interatomic distances. Complete information about molecular vibrations allows us to analyze catalytic chemical reactions.\cite{TINNEMANS20063,BRANDHORST200634,Hashimoto2019} IR spectra are used to identify functional groups and chemical bond information. However, assigning IR bands to vibrational molecular modes in the measured spectra can be difficult and requires
DFT calculations; as we mentioned earlier, the temperature is not considered in these computations
and discrepancies between experimental and computed  IR spectra can result from finite
temperature, anharmonic effects, and  multi-photon nature of experments, whereas
IR computations assume single-photon processes.\cite{Buelna,molecules26133953}
The IR spectra of isolated metal clusters in the gas phase for vanadium cluster cations
as well as for neutral and cationic niobium clusters were measured.\cite{Fielicke2005} 
Even though Cu clusters are important in catalysis and were the first clusters experimentally produced,\cite{doi:10.1021/j100211a002} the available structural information is limited to study photoelectron spectroscopy for anions,  mass spectrometry, and photodissociation spectra in the visible range.\cite{doi:10.1021/acs.jpclett.9b00539} 
Reciently, Lushchikova et al.~\cite{doi:10.1021/acs.jpclett.9b00539} determine the structure of small cationic
copper clusters based on a combination of IR spectroscopy of Cu$_n^+$-Ar$_m$  complexes and DFT calculations. In this work, the IR spectra of isomers computations were carried out using the Gaussian package under harmonic approximation at level of theory  PBPW91-D3\cite{YANAI200451}/def2TZVP  and full width at half maximum (FWHM) of 8 cm$^{-1}$ taking into account the dispersion of Grimme D3 as implemented in Gaussian code. Negative frequencies were checked in all calculations to ensure that there were not transitions states. The computed frequencies were scaled by 0.98 to estimate the observed frequencies. Here, the total IR spectrum is computed as a weighted Boltzmann sum of the single IR spectrum of each isomer of the distribution at finite temperature\cite{molecules26133953,Buelna,doi:10.1063/1.3552077,PhysRevA.70.041201} given by Equation~\ref{vcd} and the probabilities of occurrence displayed in Figure~\ref{popu}. To our knowledge, there are a few theoretical studies on IR spectra of noble metal clusters computed considering a weighted sum of the IR spectra of isomers.~\cite{PhysRevA.70.041201} Computed weighted Boltzmann IR
spectra of Cu$_{38}$ clusters at different temperatures are shown in Figure~\ref{ir2}, Appendix~\ref{appendix:b}. Notice that the transition metal clusters are very stable, and its vibrational frequencies are found below 400 cm$^{-1}$\cite{doi:10.1063/1.4822324} in good agreement with our
computed spectra displayed in Figure~\ref{ir2}, Appendix~\ref{appendix:b}. 
\subsection{UV-Visible spectra at finite temperature}
\begin{figure}[ht!]
\begin{center}  
\includegraphics[scale=0.55]{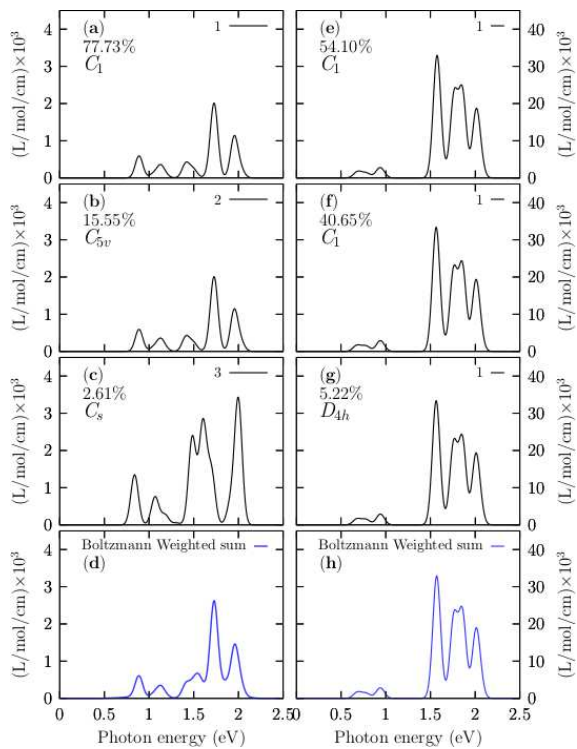}
\caption{The dependent temperature  UV-visible Boltzmann-spectra-weighted at room temperature
  of the neutral Cu$_{38}$ cluster are shown in panels (e) to (g) with Grimme's dispersion GD3 (without GD3 panels (a) to (d)).
  The computed UV-visible spectrum of each isomer is multiplied by their corresponding Boltzmann weight at finite
  temperature; then, they are summed together to produce a final Boltzmann-weighted UV spectrum.
  Each spectrum of each isomer were computed employing time-dependent
  density functional theory (TD-DFT) as implemented in Gaussian code
  at the {CAM B3LYP}/def2-SVP level of theory employing geometries optimized at B3PW91-D3/def2-SVP.
  }
\label{uv1}
\end{center}
\end{figure}
The optical properties are a source of atomic structural information, and their electronic structure determines them.\cite{ma13194300} In this paper the TD-DFT was used to compute the optical absorption spectra in the UV-visible range $0.5 < \hbar\omega <  2.5$ {eV} for the Cu$_{38}$ cluster. We employ the CAM-B3LYP functional, Def2-TZVP basis set, and 50 singlets and 50 triplets states. The transition metals clusters allow us to study the influence of the {\bf{d}} electrons on the optical properties; it is known that the {\bf{d}} electrons strongly influence the surface plasmon response.\cite{doi:10.1063/1.3552077}
We show in Figure~\ref{uv1} the UV-spectra for the Cu$_{38}$ cluster with and without the Grimme's dispersion D3. To take into account the effects of temperature on the UV-visible spectra. We consider that the UV-visible spectrum of a molecular ensemble is a weighted sum of all the individual contributions of each isomer that forms the ensemble.
 In panel labeled (e) is  displayed the UV spectra considering the  Grimme's dispersion GD3 and  for the lowest energy structure with symmetry C$_1$; this structure contribute with
 54{\%} to the total Boltzamnn spectrum. In panel (f)
 is displayed the UV-visible for the low-energy structure located at 0.16 kcal/mol at room temperature and above the putative global minimum 
 with C$_1$ symmetry, and this contributes to 40{\%} to the Boltzmann spectrum. In panel (g) is displayed the UV-visible for the low-energy structure located at 1.38 kcal/mol at room temperature and above the putative global minimum
 with D$_{4h}$ symmetry, and this contributes to 5.22{\%} to the Boltzmann spectrum. In panel (h) is displayed the Boltzmann weighted UV-Visible spectrum at room temperature. Notice that all absorption spectra,  (g) to (h), are similar; The Boltzmann weighted UV-Visible spectrum presented in Figure~\ref{uv1}, panel (h) is composed of three peaks located between  1.5 to 2 {eV} and smaller intensity peaks
 located at 0.5  and 1.0 eV. The most significant absorption peak is located at 1.6 {eV}. Our computations show five absorption peaks 
 of the  UV-visible spectrum starting at  0.5  {eV} and finishing at 2.5 {eV}. Notice that the total optical spectrum is due only to the putative global minimum; despite the number of isomers growing exponentially, the main contribution to the optical properties comes from those low energy structures very close to the global minimum where weights Boltzmann factors temperature dependents are different from zero. Interestingly, the second low-energy structure located at 0.16 kcal/mol (room temperature) above the putative lowest energy structure
 screen or blocked the contributions to the Boltzmann optical spectrum of the other structures. In this paper, we called those structures as \emph{shielding structures}. In Figure~\ref{uv1}, panels (a) to (d) are displayed the optical absorption spectra in the UV-visible range computed with structures optimized without the Grimme's dispersion D3. Notice the effects of dispersion on UV-visible spectra is introduced by the change of the relative population when the optimizations are computed without, and with dispersion,  e.g., the contribution, to the Boltzmann weighted UV-Visible spectrum, of the lowest energy structure computed without Grimme's dispersion D3 is 77{\%} meanwhile, the contribution is 54{\%} when an optimized structure is computed with Grimme's dispersion D3.
 The complete evolution of the Boltzmann weighted UV-Visible spectra for temperatures ranging from 10 to 1500 K is displayed in Figure~\ref{uv3}, Appendix~\ref{appendix:c}. 
\subsection{Molecular Dynamics}
We performed Born-Oppenheimer  \href{https://youtu.be/j0Lw8NKYUYw}{molecular dynamics (MD) of the Cu$_{38}$ cluster} employing the deMon2K program\cite{demon2k} (deMon2k v. 6.01, Cinvestav, Mexico City, Mexico, 2011) at temperature of 600 K, aiming to gain insight into the dynamical behavior of the Cu$_{38}$ cluster. The MD started from global putative minimun structure. The simulation time was 25 ps with a step size of 1 fs. All computations were performed  under scheme of ADFT,\cite{doi:10.1063/1.1771638} and for the basis set we employed the double zeta plus valence polarization (DZVP) all-electron basis. We employed the Nos\'e-Hoover thermostat to fix the temperature and the linear and angular momenta of the cluster Cu$_{38}$ were initialized to zero and conserved.
\section{Conclusions}
 For the first time, to our knowledge and from our results, we proposed the  inverted incomplete-Mackay icosahedron (IIMI)  with symmetry C$_1$
 as low-energy structure  than the incomplete-Mackay icosahedron (IMI) with symmetry C$_s$ and located 0.34 kcal/mol energy
  above the putative minimum global at 298.15 K. The yellow, the IIMI structure is the result of interchanging
  the red Cu atom depicted in the IMI structure to the position of blue atom in the IIMI structure.
  The IMI structure was reported in reference\cite{Zhang2019} as the low-energy structure.
  The HOMO-LUMO gap of the IMI structure is 0.24 eV (0.356 eV reported by previous DFT studies\cite{Zhang2019})  whereas for the HOMO-LUMO gap for IIMI structure is 0.30 eV, suggesting why the IIMI structure is energetically more stable.
 We computed the effect of symmetry on the Boltzmann populations on  Cu$_{38}$ clusters.
In spite that the number of isomers grows exponentially, the main contribution to the optical properties comes from those low energy structures
very close to the global minimum where weights Boltzmann factors temperature dependents are different from zero.
 We performed an unbiased global search for minimum energy Cu$_{38}$ clusters structures using a two-stage strategy. First a global search using a semi empirical methodology, followed by a density functional theory optimization of the best  structures from the first stage was done at different levels of theory. The temperature and entropic effects cause several competing structures because energy separation between isomers on the free energy surface is small and changes the dominant structure, so probably a mixture of isomers interconverted at temperature finite. Those energetically competing structures provide a different percentage of the entire IR spectrum. On the other hand, those higher energy structures with significant energy separation among isomers on the potential energy surface do not contribute to the entire IR spectrum. Despite that, the number of isomers grows exponentially. The main contribution to the molecular properties comes from those low-energy structures very close to the global minimum, where weight's Boltzmann factors temperature dependents are different from zero. (depends strongly on the energy separation, if the energy separation is significant, the IR spectrum going to be rigid, not changes)  
A motif is dominant in cold conditions, whereas the other motif is dominant in other hot conditions; additionally, the level of theory is decisive in the computations of TSS point dependent-temperature
Our computations clearly show (relative population) that the low-symmetry isomers become more stable at high temperatures due to the entropic effect and the fact  that energy states of molecules
follow Boltzmann distribution in all six different levels of theory.
Our unbiased global search on the free energy surface show that there is an
amorphous structure that strongly dominate at hot temperatures, as far as we know
this is a novel putative global minimum at hot temperatures.
As immediate work is the computations of the relative populations at high level of theory
and compute the $\mathcal{T}_1$ diagnostic to  determine that the computed DFT energies are not properly described by a single reference method  or contain a multireference character. So, further studies needed to be done.
\section{Acknowledgments}
C. E. B.-G. thanks Conacyt for the Ph.D. scholarship (860052). E. P.-S. thanks Conacyt for the Ph.D. scholarship (1008864).  We are grateful Universidad Polit\'ecnica de Tapachula (UPTap) for granting us access to their clusters and computational support. We are also grateful to the computational chemistry laboratory and ICEME research group at UPtap for providing computational resources, \emph{ELBAKYAN}, and \emph{PAKAL} supercomputers.
\section{Conflicts of Interest} The authors declare no conflict of interest.
\section{Funding} This research received no external funding.
\section{Abbreviations}
The following abbreviations are used in this manuscript:\\
Density Functional Theory (DFT)\\ 
Zero-Point Energy (ZPE) 
\bibliographystyle{unsrt}
\bibliography{bibliografia}
\newpage 
\appendix
\newpage
\appendix
\onecolumngrid
\section{ZPE descomposition}
\label{appendix:a}

\begin{figure}[ht!]
\begin{center}  
\includegraphics[scale=1.0]{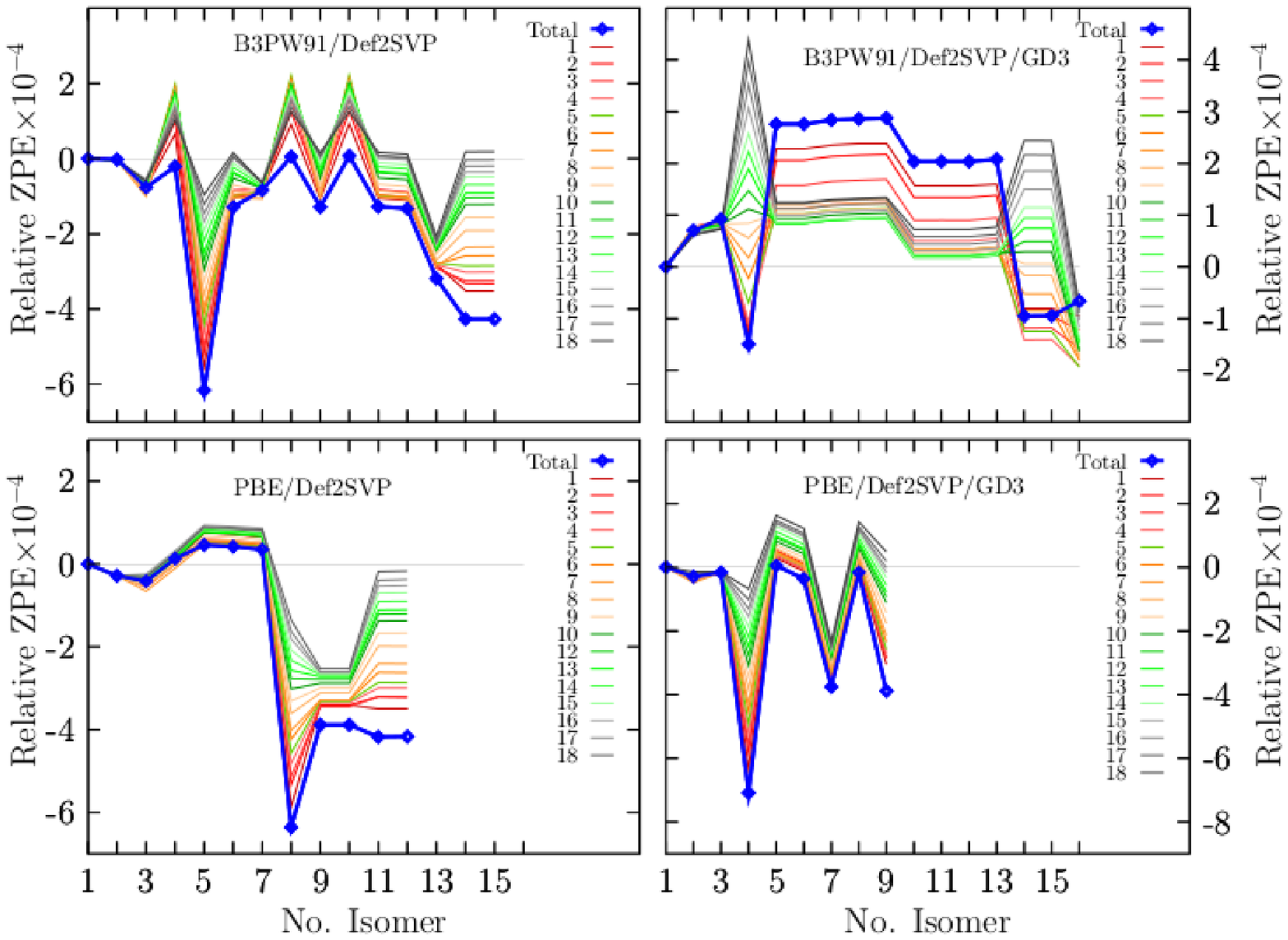}
\caption{(Color online) We show the relative zero-point energy (ZPE) decomposition as a function of the vibrational modes (Hartree/particle). In the x-axis are the isomers arranged from the lowest-energy isomer (1) to higher energy isomers (14). }
\label{popu2}
\end{center}
\end{figure}
\newpage
\section{IR spectra}
\label{appendix:b}
\begin{figure*}[ht!]
\begin{center}  
\includegraphics[scale=0.80]{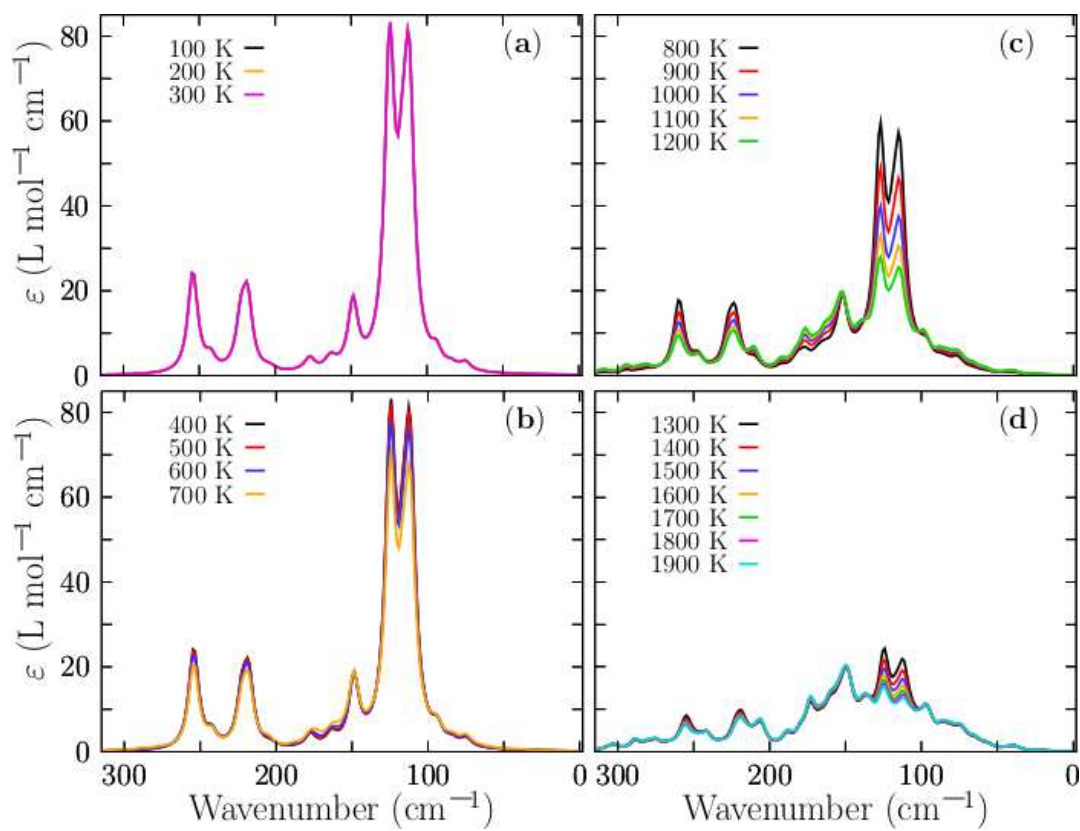}
\caption{The dependent temperature  IR Boltzmann-spectra-weighted at room temperature
  of the neutral Cu$_{38}$ cluster are shown in panels (a) to (d) for different temperatures.
  The computed IR spectrum of each isomer is multiplied by their corresponding Boltzmann weight at finite
  temperature; then, they are summed together to produce a final Boltzmann-weighted IR spectrum.
  Each spectrum of each isomer were computed employing time-dependent
  density functional theory (TD-DFT) as implemeted in Gaussian code
  at the B3PW91/def2-TZVP level of theory employing geometries optimized at B3PW91-D3/def2-SVP.
  The large change in the IR spectra occur at temperature of 824 K, as we can see in in the panel (d),
  and in good agreement with the relative occurrence displayed in Figure~\ref{popu}.}
\label{ir2}
\end{center}
\end{figure*}
\newpage 
\section{UV-Visible spectra for different temperatures}
\label{appendix:c}
\begin{figure*}[ht!]
\begin{center}  
\includegraphics[scale=1.0]{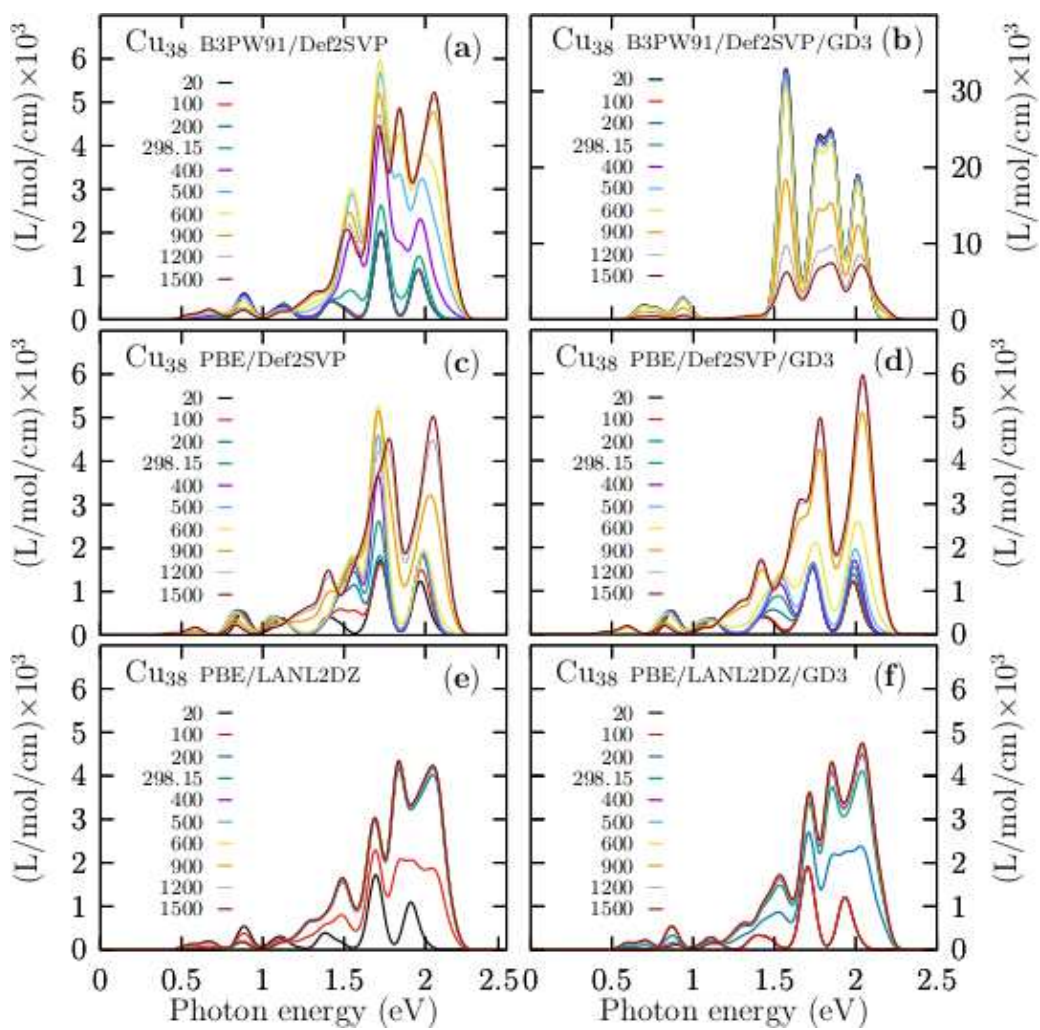}
\caption{(Color online) The dependent temperature  UV-visible Boltzmann-spectra-weighted at room temperature
  of the neutral Cu$_{38}$ cluster are shown in panels (e) to (g) with Grimme's dispersion GD3 (without GD3 panels (a) to (d)) 
  The computed UV-visible spectrum of each isomer is multiplied by their corresponding Boltzmann weight at finite
  temperature; then, they are summed together to produce a final Boltzmann-weighted IR spectrum.
  Each spectrum of each isomer were computed employing time-dependent
  density functional theory (TD-DFT) as implemented in Gaussian code
  at the {CAM B3LYP}/def2-SVP level of theory employing geometries optimized at B3PW91-D3/def2-SVP.
  }
\label{uv3}
\end{center}
\end{figure*}
\newpage
\section{Geometry at B3PW91/Def2svp level of theory}
\label{appendix:d}
\begin{figure*}[ht!]
\begin{center}  
  \includegraphics[scale=0.65]{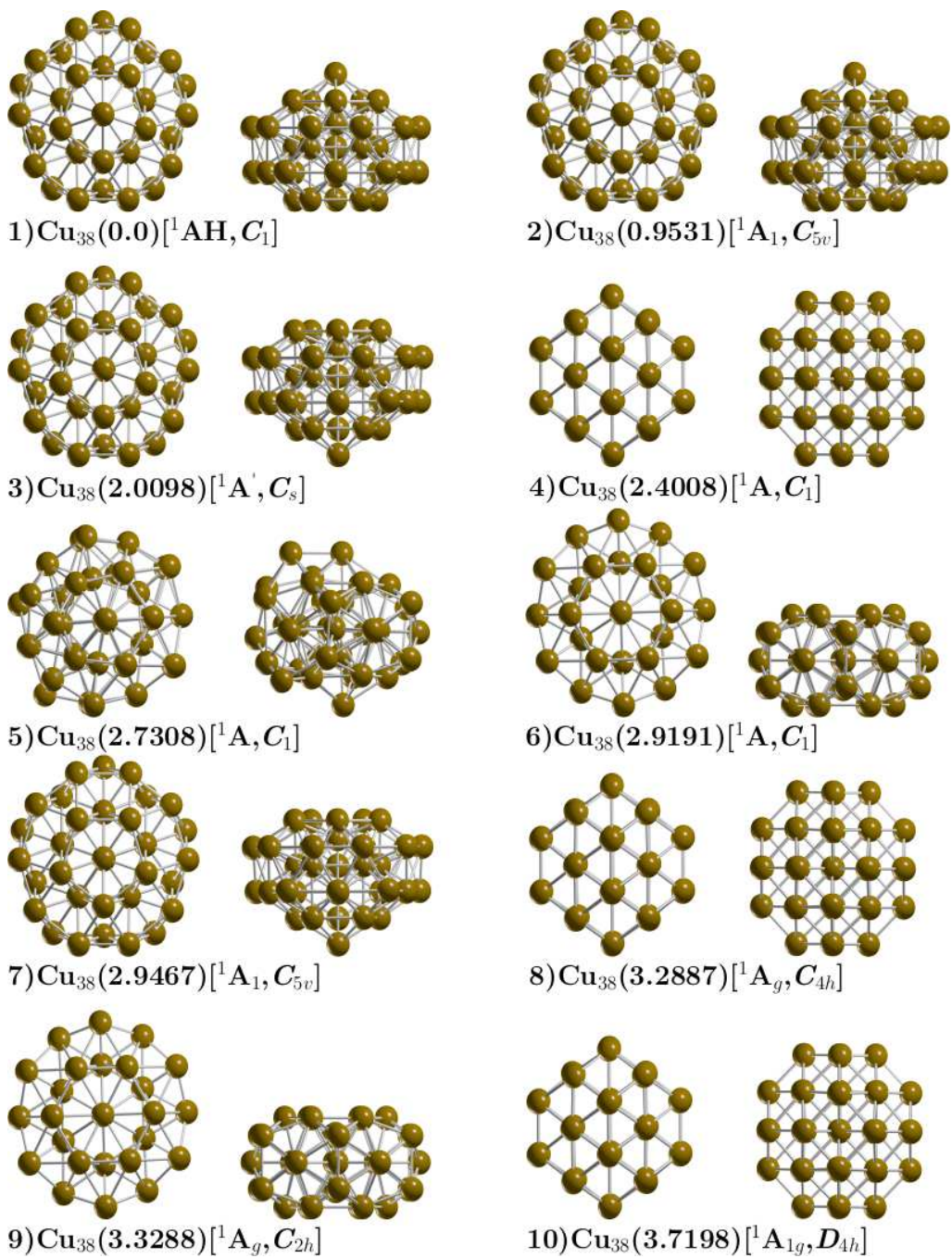}
\caption{(Color online) Optimized geometries in front and side
  views of neutral Cu$_{38}$ cluster at B3PW91/def2-SVP level of theory.
  The first letter is the isomer label,
  relative Gibbs free energies in kcal/mol (in round parenthesis) at 298.15 K,
  electronic group and group symmetry point [in square parenthesis], the probability of occurrence (in red round parenthesis) at 298.15 K.} 
\label{fig.geo}
\end{center}
\end{figure*}
\newpage
\section{XYZ atomic coordiantes}
\label{appendix:e}

\newpage
\section{XYZ atomic coordiantes at B3PW91/def2-SVP level of theory}
\label{appendix:f}
%
\typeout{get arXiv to do 4 passes: Label(s) may have changed. Rerun}
\end{document}